\documentclass[preprint,showpacs]{revtex4}
\usepackage{amsmath,amssymb,graphicx}
\usepackage{epsfig,subfigure}

\begin{document}
\author{Yamen Hamdouni}
\email{hamdouniyamen@gmail.com}
\affiliation{Department of physics, Faculty of Exact Sciences, Mentouri University, Constantine, Algeria}

\title{Aspects of the decoherence in high spin environments: Breakdown of the mean-field approximation}

\begin{abstract}
The study of the decoherence of qubits in spin systems is  almost restricted to  environments  whose constituents are  spin-$\frac{1}{2}$ particles. In this paper we consider
environments that are composed of particles of higher spin, and we investigate the consequences on the dynamics of a qubit coupled to such baths via Heisenberg
$XY$ and Ising interactions. It is shown that while the short time decay in both cases gets faster as the magnitude of the spin increases, the asymptotic behavior exhibits
an improvement of the suppression of the decoherence when the coupling is through  Heisenberg
$XY$ interactions. In the case of a transverse Ising model, we find that the mean field approximation breaks down  for high values of the spin.
\end{abstract}
\maketitle

\section{Introduction}
Modeling the  environments to which quantum systems are coupled  is a central topic in the study of the decoherence and the entanglement phenomena~\cite{prok}. Indeed,  it has become
widely accepted that the decoherence as well as the  degradation of the entanglement resource (e.g., entanglement sudden death), are solely a consequence of the interaction
of the physical systems
with their surroundings~\cite{zurek1, loss, zurek2}. This concept constitutes the main idea behind the topic of open quantum systems~\cite{petru}, which
has attracted a great deal of interest during the past decades.

The internal structure of the  environment is rather complicated in most  cases. As a matter of fact, the large number of its constituents 
makes it very difficult to deal, in an exact analytical manner, with the evolution of the  system of interest. This explains the 
reason for which one has to resort, very often, to simplified models that capture the
essential features of the environment. Needless to say that these features depend on the physical nature of the degrees of freedom characterizing its constituents.
Therefore, it  is no surprise that most of the investigations have dealt with developing various techniques that enable the elimination of 
the  spin or the bosonic degrees of freedom of the environment~\cite{kha, coish, zhang, fazio, goan, sadi, bedoor, burg, ham1,ham2, ham3, paga},
which allows one to focus  on the evolution of the central system.
The mathematical tools needed for such calculations vary depending on
whether the environment is of spin or bosonic nature. Nevertheless, the main idea behind these techniques is the same and rests in the partial trace operation, which means
that
the reduced density matrix corresponding to the central system may be obtained by taking the trace over the environmental degrees of freedom.

With the rapid progresses in the field of spintronics, it became evident that spin systems should be among the first options that have to be exploited for the implementation
of new quantum technologies~\cite{loss2,loss3}. The advantages brought by such systems reside in the fact that they can be prepared and fabricated in a scalable form that facilitates the
implementation of quantum algorithms \cite{nielsen, shor, ekert, cirac, mosca, chuang}. Therefore, it is of practical importance to consider environments that are of spin nature.
In most studies, it is usually 
assumed that the 
constituents are electrons or,  more generally,  spin-$\frac{1}{2}$ particles.

Recently, a great interest has been given to the so-called single-molecule magnets~\cite{bogani,loss4}. Common examples of such systems include  
$[{\rm Mn_3O(O_2CEt)_3(mpko)_3](ClO_4)}$ (also known as ${\rm Mn_3}$), and $[{\rm Fe_8O_2(OH)_12(tacn)_6]^{8+}}$. They are characterized by large magnetic moments, and a slow
magnetic relaxation. In addition, they exhibit configurations with large values of the spin. For example, the ground state of the single-molecule magnet
${\rm Mn_3}$ has a spin $S=6$; that corresponding 
to the ground state of $[{\rm Fe_8O_2(OH)_12(tacn)_6]^{8+}}$ is $S=10$.  Thus, it is tempting to consider the scenario in which  systems of particles of higher 
spins constitute the environment to which a central qubit is coupled. 

In this paper the emphasis is  on a particular case of spin environments, for which the whole information about the  intra-bath interactions as well as
the coupling to the central system   are encoded in the total spin operators. We have already addressed this problem  earlier  for the case of spin-$\frac{1}{2}$ 
particles~(see for example~\cite{ham2}).
 Here we shall generalize the investigation to environments formed by particles of arbitrary value of the spin. This will enable us to compare the obtained results, and to
 draw some conclusions about the advantages and the inconveniences of using this kind of environments.

The manuscript is organized as follows. Section \ref{sec1} is devoted to the study of an environment whose intra-bath interactions as well as its coupling to a central two-level
system (a qubit) is of Heisenberg $XY$ type; we derive an explicit expression for the degeneracy of the total angular momentum, and we deduce  its probability distribution 
when the size of the spin bath is sufficiently large. This is followed by the analysis of the asymptotic behavior of the state of the central system. Furthermore, we use the
Holstein-Primakoff transformation to  establish the connection between the studied spin model and a bosonic Jaynes-Cummings model. Section \ref{sec2} deals with a transverse 
Ising model; there the mean field approximation is used to linearize the problem, and we identify the critical point of the system. Then we  study analytically 
the decoherence of the qubit, and we compare the obtained results with the exact solution in the case of a vanishing transverse field. The paper is ended with a brief conclusion.

\section{Case of Heisenberg XY interactions \label{sec1}}
\subsection{Model}
The first model we shall investigate describes the coupling of a central spin-$\frac{1}{2}$ particle (a qubit) to a set
of $N$ spin-$S$ particles through Heisenberg $XY$ interactions.
The central particle is subject to  the effect of an applied
magnetic field,  the strength of which is denoted by $2\mu$. The full Hamiltonian of the system is then given by~\cite{ham2}
\begin{equation}
 H=H_S+H_{SB}+H_{B},
\end{equation}
where 
\begin{equation}
 H_S=\mu \sigma_z
\end{equation}
is the Hamiltonian of the free central spin,
\begin{equation}
 H_{SB}=
\frac{\alpha}{\sqrt{N}}\Bigl[\sigma_x\sum\limits^{N}_{i=1}{S^i_x}+\sigma_y\sum\limits^{N}_{i=1}{S^i_y}\Bigl]
\end{equation}
denotes the Hamiltonian describing the coupling of the spin to its environment, and
\begin{equation}
H_B=\frac{g}{N}\sum\limits^{N}_{i\ne j}{\Bigl(S^i_x S^j_x+S^i_y
 S^j_y\Bigl)}
 \end{equation}
 is the Hamiltonian of the spin bath. In the above, $\sigma_x$, $\sigma_y$ and $\sigma_z$ denote the usual Pauli matrices, whereas $S^i_{x,y,z}$ are the components of the 
 spin operator of the i'th spin in the environment. The constants $\alpha$ and $g$ are the coupling strengths, and are assumed positive.
 
 It is worth noting that the case for which the environment is composed of particles with spin -$\frac{1}{2}$ has been thoroughly investigated in Ref.~\onlinecite{ham2}. 
 There, the rescaling of the coupling constants $\alpha$ and $g$ by the factors $\sqrt{N}$ and $N$, respectively,  enabled us to discuss the case when $N$ is infinite. From a statistical point of view, the above 
 rescaling ensures the well behavior of the free energy of the spin bath. In this work, we shall follow a slightly different approach to deal with
 the case of a large number of environmental spins, namely, we shall derive a probability distribution for the total spin angular momentum, that can   be  used for large
 but finite number of spins within the  environment.
 
 To fully describe the state of the qubit, we need to determine the evolution in time of its density matrix $\rho$. Usually, one assumes that the qubit is initially uncorrelated
 with its environment, and that the latter is in thermal equilibrium at temperature $T$. Since the evolution of the total system is unitary, it is sufficient to eliminate
 the environmental degrees of freedom from the evolution equation by tracing out the states of the bath.
 In our case it is more convenient to work in the basis composed of the common eigenvactors of the operators
 $J^2$ and $J_z$, where 
 \begin{equation}
 \vec J=\sum\limits_{i=1}^N \vec{S^i}.
\end{equation}
These vectors are denoted by $|j,m\rangle$ such that (we assume $\hbar=1$)
\begin{equation}
 J^2|j,m\rangle=j(j+1)|j,m\rangle,  \qquad  J_z|j,m\rangle=m|j,m\rangle.
\end{equation}

 The evolution in time of the density matrix of the system we are interested in is thus given by  [note that $\beta=1/(k_B T$), $k_B$ being Boltzmann's constant]
 \begin{widetext}
 \begin{equation}
  \rho(t)=\frac{1}{Z}\sum\limits_{j,m} \nu(j,N;S) \langle j,m|\exp(-iH t)\rho_S(0)\otimes\exp(-\beta H_B)\exp(iH t)|j,m \rangle \label{red}.
 \end{equation}
\end{widetext}
The quantity $\nu(j,N;S)$ stands for the degeneracy corresponding to the value $j$ of the total spin angular momentum. Knowing the degeneracy, one can 
decompose the total spin space of the environment as follows:
\begin{equation}
 {\mathbb C^{ 2S+1}}^{\otimes N}=\bigoplus\limits_{j}^{S N} \nu(j,N;S) {\mathbb C^{ 2j+1}},
\end{equation}
 where $\mathbb C $ is the field of complex numbers, and the sum over $j$ runs from 0 ($\tfrac{1}{2}$) to $N S$, when $N S$ is even (odd).
 Hence our next task resides in the determination of
the degeneracy $\nu$, which we fulfill in the next subsection.

\subsection{The degeneracy and the distribution of the quantum number $j$}

In order to find the expression of the degeneracy $\nu$, we introduce the spaces:
\begin{equation}
 F_m= \{  \mathcal{V}_{jm}\in {\mathbb C^{ 2S+1}}^{\otimes N},\quad  J_z  \mathcal{V}_{jm}=m  \mathcal{V}_{jm} \},
\end{equation}
and
\begin{eqnarray}
 E_{j,m}=  \{ \mathcal{V}_{jm}\in {\mathbb C^{ 2S+1}}^{\otimes N},\ J_z  \mathcal{V}_{jm}=m  \mathcal{V}_{jm},\nonumber \\
 \quad J^2  \mathcal{V}_{jm}=j(j+1)  \mathcal{V}_{jm} \}.
\end{eqnarray}
It can easily be verified that the dimension of the space $F_m$ is given by
\begin{widetext}
\begin{equation}
 {\rm dim} F_m=\sum\limits_{L_{-S},L_{-S+1},\cdots, L_{S}}^N\frac{N!}{\prod\limits_{k=0}^{2S} (L_{-S+k})!} \delta\Biggl(\sum\limits_{\rho=0}^{2S}L_{-S+\rho}
 ,N\Biggl)\delta\Biggl(\sum\limits_{\rho=0}^{2S}(S-\rho)L_{-S+\rho},m\Biggl).\label{dimem}
\end{equation}
\end{widetext}
In the above equation, the quantities  $L_{i}$ (with $i=-S,-S+1,-S+2,\cdots S-1,S $) are integer numbers taking on values from 0 to $N$. The expression~(\ref{dimem}) can
be further simplified to the form
\begin{widetext}
\begin{eqnarray}
 {\rm dim} F_m & =& \sum\limits_{L_{-S+2},L_{-S+3},\cdots, L_{S}}^N\frac{N!}{\prod\limits_{k=2}^{2S} (L_{-S+k})!} 
 \nonumber \\ &\times& \Biggl[\Biggr(SN+m-\sum\limits_{\rho=2}^{2S}\rho L_{-S+\rho}
 \Biggr)!\Biggr((1-S)N-m+\sum\limits_{\rho=2}^{2S}(\rho-1) L_{-S+\rho}
 \Biggr)!\Biggr]^{-1}\label{dimem2}.
\end{eqnarray}
\end{widetext}

Now since 
\begin{equation}
 F_m=\bigoplus\limits_{j=m}^{NS}E_{j,m},
\end{equation}
which means that
\begin{equation}
 {\rm dim} E_{j,m}= {\rm dim}F_{j}- {\rm dim}F_{j+1}=\nu(j,N;S),
\end{equation}
we obtain after some algebra
\begin{widetext}
\begin{eqnarray}
 \nu(j,N;S)=\sum\limits_{L_{-S+2},L_{-S+3},\cdots, L_{S}=0}^N\frac{N!}{\prod\limits_{k=2}^{2S} (L_{-S+k})!} \Biggl(\frac{(2S-1)N+2j+1-
 \sum\limits_{\rho=2}^{2S}(2\rho-1)L_{-s+\rho}}
 {SN+j+1-\sum\limits_{\rho=2}^{2S}\rho \ L_{-s+\rho}}\Biggr)\nonumber \\
 \times \Biggl[\Bigl(SN+j-\sum\limits_{\rho=2}^{2S}\rho \ L_{-s+\rho}\Bigr)!\Bigl((1-S)N-j+\sum\limits_{\rho=2}^{2S}(\rho-1) L_{-s+\rho}\Bigr)!\Biggr]^{-1}\label{deg}.
\end{eqnarray}
\end{widetext}
Notice that in the particular case where  $S=1/2$, the degeneracy simplifies to $\binom{N}{N/2-j}-\binom {N}{N/2-j-1}$.~\cite{vonder}

In this way we can assign to the quantum number $j$  a probability distribution, which we designate by $P(j)$,  as follows:
\begin{equation}
 P(j)=\frac{2j+1}{(2S+1)^N}\nu(j,N;S).
\end{equation}
This distribution corresponds to a tracial state of a randomly distributed set of $N$  independent spin-$S$ particles. It allows, under convergence conditions, for the calculation of the expectation value  of
any quantity that 
depends on the total angular momentum number $j$; the latter has to be dealt with as a continued real random variable when $N$ is sufficiently large.

Before we proceed further, let us notice that equation (\ref{deg})  implies that whatever the values of $N$ and $S$ are, we have:
\begin{equation}
 \nu(NS,N;S)=1, \quad \nu(NS-1,N;S)=N.
\end{equation}

Furthermore, since
\begin{eqnarray}
 \mathbb{C}^{(2S+1)\otimes N}\otimes \mathbb{C}^{2S+1}&&=\bigoplus_j^{NS}\nu(j,N;S)\mathbb{C}^{(2j+1)}\otimes \mathbb{C}^{2S+1}\nonumber \\
 &&=\bigoplus_j^{NS}\nu(j,N;S)\bigoplus_{j^{'}=|j-S|}^{j+S}\mathbb{C}^{(2j'+1)},
\end{eqnarray}
it follows that
\begin{equation}
 \nu(j,N+1;S)=\sum\limits_{j'=|j-S|}^{j'+S}\nu(j',N;S).
\end{equation}
Actually, the above equality is a special case of the property~\cite{ham3}
\begin{eqnarray}
 \nu(J,N_1+N_2;S)&=&\frac{1}{2J+1} \sum_{j_1,j_2,m_1,m_2,M} \nu(j_1,N_1;S)\nu(j_2,N_2;S ) \nonumber \\
 &\times &\langle j_1m_1j_2m_2|JM\rangle^2.
\end{eqnarray}

When $N$ becomes very large, the probability distribution $P(j)$ approaches a Gaussian distribution, the form of which may be inferred  from the expression of $\nu$ together
with the fact that 
\begin{eqnarray}
 \frac{1}{(2S+1)^N}\mathrm{tr}\biggl(\sum\limits_{k,j=1}^{N}\vec{S}^k \vec{S}^j\biggr)&=& \frac{3}{(2S+1)^N}\mathrm{tr}\biggl(\sum\limits_{k,j=1}^{N}S_{z}^k S_{z}^j\biggr) \nonumber\\
  &=&NS(S+1).
 \label{trr}
\end{eqnarray}
Taking into account equations (\ref{deg}) and (\ref{trr}), we find that the explicit expression of the distribution $P(j)$ takes the form
\begin{equation}
 P(j)=\frac{6j^2}{NS(S+1)} \sqrt{\frac{3}{2\pi NS(S+1)}}\exp\biggl(-\frac{3j^2}{2S(S+1)N}\biggr).\label{prob}
\end{equation}
Hence given a function $f$ of the random variable $j$, its mean value can be evaluated as:
\begin{equation}
 \langle f(j) \rangle=\int\limits_0^{\infty} P(j) f(j) dj.
\end{equation}
In particular, we find that the mean value of $j$ reads
\begin{equation}
 \langle j \rangle= 2 \sqrt{\frac{2}{3\pi}}\sqrt{NS(S+1)},
\end{equation}
which leads to
\begin{equation}
 (\Delta j)^2=\langle j^2 \rangle-\langle j \rangle^2=\biggl(1-\frac{8}{3\pi}\biggr) NS(S+1).
\end{equation}
This shows that the width of the distribution is proportional to $\sqrt{N}$, which explains, once more, the reason for which the coupling constants
$g$ and $\alpha$ have been 
rescaled by the respective powers of the size of the  environment. As we have mentioned above the rescaling ensures that the Helmholtz free energy of the system is extensive.
This is necessary in order to study what is referred to as the thermodynamic limit (more precisely, the limit $N\to \infty$). For example,
the order of magnitude 
of $N$ is $10^6$ in a quantum dot. On the other hand, for the known molecules, the order of magnitude of the spin is $S\sim 10$. Note, nevertheless, that even for 
the case where $S\sim 3$, one is
actually dealing
with a large spin, for which the quasiclassical approximation may be employed~\cite{das}. In the latter approximation, the environmental  spin is dealt
with as a classical vector. It is also possible to use spin coherent states for large $S$ to express the Hamiltonian classically, and to find the quantum corrections to the latter
in the form of a series of powers of $1/S$~\cite{garanin}. The present work  deals with the environmental  spins quantum mechanically.

Another point worth mentioning, relative to the large $S$ limit, is the connection with a bosonic environment. This link can be established using the
Holstein-Primakoff transformation which maps the spin operators  $S^k_\pm$  to bosonic ones as follows:~\cite{holstein}
\begin{equation}
 S^k_-=\sqrt{2S}\sqrt{1-\tfrac{a_k^\dag a_k}{2S}} a_k, \quad  S^k_+=\sqrt{2S}a^\dag_k\sqrt{1-\tfrac{a_k^\dag a_k}{2S}},
\end{equation}
with $[a_k, a_{k'}^\dag]=\delta_{kk'}$. When $S>>1$, we may write as a first approximation
\begin{equation}
 S_-^k\approx \sqrt{2S} a_k, \quad S_+^k\approx   \sqrt{2S} a^\dag_k.
\end{equation}
Let us introduce the operators
\begin{eqnarray}
 B&=&\frac{1}{\sqrt{N}}\sum_{k=1}^N a_k,\\
 B^\dag&=&\frac{1}{\sqrt{N}}\sum_{k=1}^N a^\dag.
\end{eqnarray}
It is easily verified that these operators satisfy $[B, B^\dag]=1$, that is they are also bosonic operators. Then, to a good approximation, we can write
in the limit $N\to\infty$ 
\begin{eqnarray}
 H_{SB}&=& 2\alpha\sqrt{2S}(\sigma_- B^\dag+\sigma_+ B),\label{bose1}\\
 H_B&=&2gSB^\dag B.\label{bose2}
\end{eqnarray}
This shows that under the above assumptions, the model is equivalent to a Jaynes-Cummings model~\cite{jaynes}. We see also that the coupling constants of the new Hamiltonian
are proportional to the magnitude of the spin, which is analogous to the results obtained in the context of the spin wave theory~\cite{majlis}. (See Appendix~\ref{app2}
for more details about the  time
evolution of this bosonic model.)

\subsection{Time evolution}

The time evolution of the central qubit can be studied  in the large $N$ limit using the time evolution operator that has been derived in Ref.\onlinecite{ham2}.
Using that form, the elements of the reduced 
density matrix may be calculated by virtue of equation~(\ref{red}). The trace operation over the environmental degrees of freedom is
carried out using the probability 
distribution of $j$ provided that $N$ is sufficiently large. This considerably facilitates the calculation, since the direct use of the degeneracy $\nu$ requires the 
evaluation of sums of terms that grow rapidly with $N$ and $j$. Furthermore, the analytical form of $P$ 
makes it possible to find in close analytical form the expression of the asymptotic reduced density matrix (see bellow).

The only difficulty we have to face  here rests 
in the fact that the quantum number $m$ should be dealt with as a random
variable that is dependent on $j$. To overcome this difficulty, we adopt, as a first step, the approximation in which $m^2$  is replaced by $\theta  j^2 $ where 
$\theta$ is a yet-to-be-determined parameter. Then, for sufficiently large values of $N$ we find, for instance, that
\begin{widetext}
\begin{eqnarray}
 \rho_{12}(t)&=&\rho_{21}^*(t) =\frac{\rho_{12}(0)}{\mathcal Z}   \int\limits_0^\infty P(j) e^{-\frac{g\beta (1-\theta) j^2}{N}} \biggl \{\cos^2(t \sqrt{\mu^2+(1-\theta)\alpha^2 j^2/N})
 \nonumber \\
 &-& \dfrac{\mu^2}{\mu^2+\alpha^2 (1-\theta)j^2/N} \sin^2(t \sqrt{\mu^2+(1-\theta)\alpha^2 j^2/N}) \nonumber \\
 &+& \dfrac{i \mu}{\sqrt{\mu^2+\alpha^2 (1-\theta)j^2/N}} \sin(2t \sqrt{\mu^2+(1-\theta)\alpha^2 j^2/N})\biggr \}dj,
\end{eqnarray}
\end{widetext}
where
\begin{equation}
 \mathcal Z = \biggl[1+2 S(S+1)g \beta (1-\theta)/3\biggr]^{-3/2}.
\end{equation}

  Similarly, the diagonal elements of the reduced density matrix are calculated as:
  \begin{widetext}
\begin{eqnarray}
&&  \rho_{11}(t)=1-\rho_{22}(t)= \frac{\rho_{11}(0)}{\mathcal Z}  \int\limits_0^\infty P(j) e^{-\frac{g\beta (1-\theta) j^2}{N}}
\biggl \{\cos^2(t \sqrt{\mu^2+(1-\theta)\alpha^2 j^2/N})\nonumber \\
 && + \dfrac{\mu^2}{\mu^2+\alpha^2 (1-\theta)j^2/N} \sin^2(t \sqrt{\mu^2+(1-\theta)\alpha^2 j^2/N}) \biggr \}dj +\frac{\rho_{22}(0)}{\mathcal Z} 
 \int\limits_0^\infty P(j) e^{-\frac{g\beta (1-\theta) j^2}{N}}  \nonumber \\
&& \times \biggl \{\dfrac{\alpha^2/N}{\mu^2+\alpha^2 (1-\theta)j^2/N} \sin^2(t \sqrt{\mu^2+(1-\theta)\alpha^2 j^2/N})\biggr \}dj.
\end{eqnarray}
\end{widetext}

Having determined the  analytical forms of the elements of $\rho$ , it is now possible to deduce their asymptotic values
by making use of the Riemann-Lebesgue lemma. This yields
\begin{widetext}
\begin{eqnarray}
 \psi= \lim_{t\to \infty} \rho_{12}(t)/\rho_{12}(0)&&=\frac{1}{2}-\biggl(\frac{\mu}{\alpha} \biggr)^2\biggl[\beta g +\frac{3}{2(1-\theta) S(S+1)}\biggr] \nonumber \\
  &&-\sqrt{\pi}\biggl(\frac{\mu}{\alpha} \biggr)^3\biggl[\beta g +\frac{3}{2(1-\theta) S(S+1)}\biggr]^{3/2}\nonumber \\ 
  && \times  \exp\Biggl[(\mu/\alpha)^2 \biggl(\beta g +\frac{3}{2(1-\theta) S(S+1)}\Biggr)\Biggr]\nonumber \\ 
  && \times {\rm erfc}\Biggl[ \frac{\mu}{\alpha}\sqrt{ \biggl(\beta g +\frac{3}{2(1-\theta) S(S+1)}\Biggr)}\Biggr],\label{man}
\end{eqnarray}
\end{widetext}
where ${\rm erfc}(x)$ designates the complementary error function, and  we have used the symbol $\psi$ for later convenience and ease of notation. One can see that the right-hand side of the latter equation is independent of the number of environmental
spins, which is a direct consequence of the rescaling of the coupling strengths $g$ and $\alpha$. 

Comparing the latter result with that obtained for $S=1/2$, we come to the conclusion that the parameter $\theta$ should vanish, i.e $\theta\equiv 0$. In order to explain
this result, it suffices to notice that the interaction between the bath's spins is of Heisenberg $XY$ type  whose form includes only the $x$ and $y$ components of the spin operators. 
The spin coupling makes it more probable for the total spin vector to lie within the $x$-$y$ plane. 
Hence it is plausible to neglect $m^2$ compared to $j^2$ which, obviously, contains the contribution of the three components of the total spin operator $\vec J$. We only
need to put $\theta=0$ into equation~(\ref{man}) in order to determine the long-time behavior of the off-diagonal element $\rho_{12}$; that corresponding to
the diagonal element $\rho_{11}$ reads
\begin{widetext}
\begin{eqnarray}
  \lim_{t\to \infty} \rho_{11}(t)&=&\rho_{11}(0)\biggl\{\frac{1}{2}+\biggl(\frac{\mu}{\alpha} \biggr)^2\biggl[\beta g +\frac{3}{2  S(S+1)}\biggr] \nonumber \\
  &+&\sqrt{\pi}\biggl(\frac{\mu}{\alpha} \biggr)^3\biggl[\beta g +\frac{3}{2  S(S+1)}\biggr]^{3/2}  \exp\Biggl[(\mu/\alpha)^2 \biggl(\beta g +\frac{3}{2  S(S+1)}\Biggr)\Biggr]\nonumber \\ 
  &\times & {\rm erfc}\Biggl[ \frac{\mu}{\alpha}\sqrt{ \biggl(\beta g +\frac{3}{2  S(S+1)}\Biggr)}\Biggr]\Biggr \}
  +\rho_{22}(0) \Biggl\{\sqrt{\pi}( \mu /\alpha)^3 \biggl[\beta g +\frac{3}{2  S(S+1)}\biggr]^{3/2}\nonumber \\ 
  &\times & \exp\Biggl[(\mu/\alpha)^2 \biggl(\beta g +\frac{3}{2  S(S+1)}\Biggr)\Biggr]{\rm erfc}
  \Biggl[ \frac{\mu}{\alpha}\sqrt{ \biggl(\beta g +\frac{3}{2  S(S+1)}\Biggr)}\Biggr]\Biggr \}.
\end{eqnarray} 
\end{widetext}
\begin{figure}[htb!]
  {\centering
\resizebox*{0.47\textwidth}{!}{\includegraphics{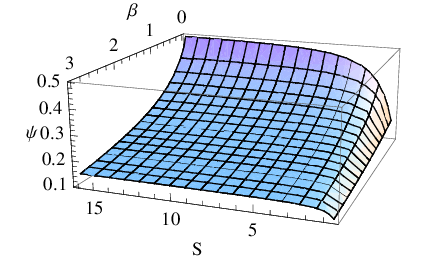}} \par}
  \caption{(Color online) The asymptotic value of the coherence $\psi$ for different values of $S$ and $T$. The remaining  parameters are $g=1$, $\mu=\alpha$.}
 \label{fig1}
\end{figure}

\begin{figure}[htb!]
 {\centering
 \resizebox*{0.43\textwidth}{!}{\includegraphics{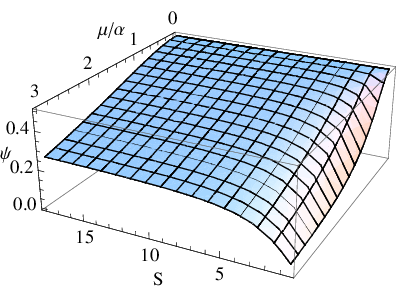}} \par}
  \caption{(Color Online) The asymptotic value of the coherence $\psi$ for different values of $S$ and $\mu/\alpha$. The other parameters are $g=1$, $\beta=0.1$ .}
 \label{fig2}
\end{figure}

 We have displayed in Figures~\ref{fig1} and~\ref{fig2}
the variation of the asymptotic value of the off-diagonal element $\rho_{12}$   as a
function of $S$, $\beta$ and  $\mu/\alpha$ for $g=1$. It is clear that the above matrix element assumes larger values as $S$ increases, indicating that the partial suppression of
the decoherence may be 
improved in environments whose constituents are of  high spin. This effect becomes more apparent as the temperature increases.

From a statistical point of view, the above results imply that the amount of information accessible at long times to the central system, which initially has been leaked to the 
spin bath, is greater when the magnitude of $S$ increases. Indeed, initially the qubit looses rapidly its coherence due to the coupling to the environment; afterwards, as things
randomize, the qubit has the tendency to adhere  the  state that minimizes the loss of coherence. The number of accessible  bath's state vectors is equal to $(2S+1)^N$. 
Clearly this number increases as $S$ increases, allowing more options for the central system to  select among the above states, 
those which display the minimal decoherence due to the coupling to the bath. This process cannot go on indefinitely, since a saturation does
emerge as the magnitude of the spin
$S$ increases; indeed, when $S>>1$, we find that
\begin{eqnarray}
\rho_{12}(\infty)/\rho_{12}(0)&=&\frac{1}{2}-\biggl(\frac{\mu}{\alpha} \biggr)^2\beta g -\sqrt{\pi}\biggl(\frac{\mu}{\alpha} \biggr)^3(\beta g)^{3/2}\nonumber \\ 
  && \times  \exp[(\mu/\alpha)^2 \beta g]
   {\rm erfc}\Bigl[ \frac{\mu}{\alpha}\sqrt{ \beta g }\Bigr].
\end{eqnarray}

At short times, the decay of the reduced density matrix is purely Gaussian, which is typical for the non-Markovian dynamics.
For example, the evolution of the off-diagonal element can be approximated at short times by
\begin{equation}
 |\rho_{12}(t)|=|\rho_{12}(0)|e^{-\frac{t^2}{\tau_D^2}}
\end{equation}
where the decoherence time $\tau_D$ can be determined via the second-order master equation, describing the evolution of the open system (see, e.g., \cite{ham3}).
Explicitly, we find that
\begin{equation}
 \tau_D=\frac{1}{\alpha} \sqrt{\beta g+\frac{3}{2S(S+1)}}.
\end{equation}
From the latter expression, we deduce that the larger $S$, the shorter $\tau_D$, meaning that the decoherence is faster when $S$ is large. The minimum value of $\tau_D$ is 
given by
\begin{equation}
 \tau_D^{\rm min}=\frac{1}{\alpha} \sqrt{\beta g}.
\end{equation}
Hence, as far as the discussion is concerned with the magnitude of $S$,  there exits an apparent competition between the speed of
the loss of coherence of the central system at short times and its asymptotic state. To be more specific, we note that, while the
decoherence time constant takes its smallest  values for high spins, the partial recovery of quantum interferences gets better, and {\it vice versa}.
This means that if one is interested in the short time behavior, it is much better to use a spin bath
that is composed of spin-$\frac{1}{2}$ particles, which yields the largest values of $\tau_D$. On the contrary, if the application requires the optimal state
at long times, then it is more convenient to use a spin environment with $S$ sufficiently large.

\section{Case of transverse Ising model \label{sec2}}
\subsection{Model}
The second model we shall investigate is a generalization of that studied by Lucamarini {\it et al} \cite{paga}. Here we would like to describe the interaction of a qubit with
a ferromagnetic symmetry-broken spin environment that  is
composed of $N$ spin-$S$ particles; the latter are coupled to each other via Ising type interactions and are subject to a transverse magnetic field along the $x$ direction.
The Hamiltonian of the total system is given by
\begin{equation}
 H=H_0+H_{SB}+H_B
\end{equation}
with
\begin{eqnarray}
 H_0&=&\mu S^0_z, \\
 H_{SB}&=&-\frac{J_0}{\sqrt{N}}S^0_z\sum\limits_{i=1}^N S^i_z,  \\
 H_B&=&-w \sum\limits_{i=1}^N S^i_x-\frac{J}{N}\sum\limits_{i, j}^N S^i_zS^j_z . 
\end{eqnarray}
Notice that the coupling constant of the qubit to the environment has been denoted by $J_0$, whereas the strength of the long range intra-bath interactions
has been designated by $J$. Moreover, $\mu$ and $w$ are the strengths of the applied magnetic fields.

The bath's Hamiltonian can be linearized using the mean field approximation; afterwards,  the problem may be fully studied analytically as we shall see bellow. Indeed, 
the mean field
approximation yields the following form of the Hamiltonian
\begin{equation}
 H_B^{mf}=-w \sum\limits_{i=1}^N S^i_x-2Jm\sum\limits_{i=1}^N S^i_z+N J m^2,
\end{equation}
where $m$ is the order parameter of the phase transition, the value of which can be fixed by the self-consistency condition that arises from minimizing the free energy  $F=-1/(N\beta) \ln Z_N$. 
In the avove, the partition function corresponding to the mean-field
approximation Hamiltonian $H_B^{mf}$ is given by
\begin{equation}
 Z_N={\rm tr} e^{-\beta H_B^{mf}}.
\end{equation}
The value of $m$ runs from  $S$ to 0, provided that  the temperature  varies within the interval $0\leq T \leq T_c$, where $T_c$ is a critical temperature that will be derived 
shortly.

It can be shown by a suitable rotation within the $x$-$z$ plane that for  $S$ even ( the case $S$ odd yields the same results):
\begin{widetext}
\begin{eqnarray}
 Z_N=e^{-\beta m^2 J N} \prod\limits_{k=1}^N \Biggl[1+{\rm tr } \bigoplus\limits_{\ell=1}^S \biggl \{ \cosh(\beta \ell \tan(\phi)) {\mathbb I_2}+
 \frac{\sinh(\ell \tan(\phi))}{\beta \tan(\phi)}\sigma_z \biggr \}\Biggr ],
\end{eqnarray}
\end{widetext}
where ${\mathbb I_2}$ refers to the two-dimensional unit matrix, and $\phi$ is given by
\begin{equation}
 \tan(\phi)=\sqrt{w^2+4 J^2 m^2}:=\Theta.
\end{equation}
It follows that 
\begin{equation}
 Z_N=e^{-\beta m^2 J N} \biggl[1+2 \sum\limits_{\ell =1}^S \cosh\bigl(\ell \beta \Theta \bigr) \biggr]^N.
\end{equation}
To evaluate the above expression, it suffices to use the exponential form of the {\rm cosh } function; then one gets two sums involving geometric series, that can easily be
calculated to yield
\begin{equation}
 Z_N=e^{-\beta m^2 J N}\Biggl[1+2 \cosh\Bigl[\Bigl(\frac{S+1}{2}\Bigr)\beta \Theta\Bigr] \frac{\sinh(S\beta\Theta/2)}{\sinh(\beta\Theta/2)}\Biggr]^N.
\end{equation}
By minimizing the free energy, we obtain the following self-consistency equation
\begin{equation}
 \frac{\Theta}{J}=\frac{S \sinh\Bigl(\beta (S+1)\Theta\Bigl)-(S+1)\sinh\Bigl(\beta S \Theta\Bigr)}{\sinh\Bigl(\beta  \Theta/2\Bigr)\sinh\Bigl(\beta (2S+1) \Theta/2\Bigr)}.
\end{equation}
As a consequence, the critical point at which the phase transition occurs is given by
\begin{equation}
 T_c=\frac{2JS(S+1)}{3k_B}.\label{cric}
\end{equation}
Remarkably, this result, which is based on the mean field approximation, agrees with the probability distribution $P(j)$, shown in equation~(\ref{prob}). Indeed, to
ensure the convergence of the expectation value of the bath's function $\exp\{J\beta j^2/N\}$, we should have $ J\beta /N < 3/(2N S(S+1))$, which yields exactly the same critical
temperature as~(\ref{cric}). A similar expression  of the critical temperature has been derived  in~Ref.~\onlinecite{kittel}; a close one is reported  
 in~Ref.~\onlinecite{das} where  the mean field theory and semiclassical arguments are used. This dependence on the spin is obvious since the size of the spin space
 increases, leading to an  enlargement  of the ordered phase, and hence an increase of the critical temperature.

The ordered phase corresponds to the temperature interval $0\le T \le T_c$; in this phase, the supplementary  condition
\begin{equation}
 \frac{w}{J}<\frac{S \sinh\Bigl(\beta (S+1)w\Bigl)-(S+1)\sinh\Bigl(\beta S w \Bigr)}{\sinh\Bigl(\beta w /2\Bigr)\sinh\Bigl(\beta (2S+1)w /2\Bigr)}
\end{equation}
is satisfied.

\subsection{Coherence evolution}
Without loss of generality, we  suppose that $S=1$; this particular case captures the main features of the dynamics, and is relatively simple to study analytically.
Larger values of $S$ show the same behavior, but are a bit more complicated since the mathematical formulas are more cumbersome (see  Appendix~\ref{app1} for $S=\frac{3}{2}$ 
and $S=2$). The  physical
conclusions drawn from all these cases  are the same.

If the spin bath is in thermal equilibrium at temperature $T$, then under the mean-field approximation,
 the evolution in time of the reduced density matrix of the central qubit is described by
 \begin{widetext}
 \begin{eqnarray}
  \rho(t)&=&{\rm tr_B}\sum_{\ell,n}\frac{\rho_{\ell n}(0)}{Z_N} \exp\Biggl\{-it\Bigl(\mu S_z^0-\frac{J_0}{\sqrt{N}}S^0_z\sum\limits_{i=1}^N S^i_z 
  -w \sum\limits_{i=1}^N S^i_x-2Jm\sum\limits_{i=1}^N S^i_z+N J m^2\Bigr)\Biggr\}\nonumber \\
  && \times |\ell\rangle\langle n| \exp\Biggl\{-\beta \Bigl(  -w \sum\limits_{i=1}^N S^i_x-2Jm\sum\limits_{i=1}^N S^i_z+N J m^2\Bigr)\Biggr\}\nonumber \\
  && \times \exp\Biggl\{it \Bigl(\mu S_z^0-\frac{J_0}{\sqrt{N}}S^0_z\sum\limits_{i=1}^N S^i_z 
  -w \sum\limits_{i=1}^N S^i_x-2Jm\sum\limits_{i=1}^N S^i_z+N J m^2\Bigr)\Biggr\},
 \end{eqnarray}
 \end{widetext}
 where $\ell,n\equiv \pm1$, and $S_z^0|\ell\rangle=\ell/2|\ell\rangle$. In particular: 
\begin{eqnarray}
 \rho_{12}(t)\equiv \rho_{+-}(t)=\rho_{+-}(0) g(t),
\end{eqnarray}
where the function $g$ describes the decoherence of the qubit and is defined by
\begin{widetext}
\begin{eqnarray}
 g(t)&=&\frac{1}{[1+2\cosh(\beta \Theta)]^N} \prod\limits_{k=1}^N{\rm tr}\exp\Biggl\{it\Biggl[\Bigl(\frac{J_0}{2\sqrt{N}}+2Jm\Bigr)S_z^k
 +wS_x^k)\Biggr]\Biggl\} \nonumber \\
 &\times & \exp\Bigl\{\beta (w S_x^k+2Jm S_z^k)\Bigr\} \exp\Biggl\{it\Biggl[\Bigl(\frac{J_0}{2\sqrt{N}}-2Jm\Bigr)S_z^k
 -wS_x^k)\Biggr]\Biggl\} \label{gon}.
\end{eqnarray}
\end{widetext}
We see that the magnetic field $\mu$ does not affect the off-diagonal elements of the reduced density function; this is the reason for which we neglect it in the subsequent
discussion.

The  function $g(t)$ can be calculated as follows. First let us introduce the operator
\begin{equation}
 G=\alpha S_z + \gamma S_x,
\end{equation}
where $\alpha$, $\gamma$ are complex numbers, and
\begin{equation}
 S_x=\begin{pmatrix}
      0 && \frac{1}{\sqrt{2}} && 0 \\
      \frac{1}{\sqrt{2}} && 0 && \frac{1}{\sqrt{2}}\\
      0 && \frac{1}{\sqrt{2}} && 0
     \end{pmatrix},  \qquad  S_z=\begin{pmatrix}
      1 && 0 && 0 \\
      0 && 0 && 0\\
      0 && 0 && -1
     \end{pmatrix}.
\end{equation}
 Then, it can be shown by induction  that  for  $k\neq 0$,
\begin{equation}
 G^{2k}= (\alpha^2+\gamma^2)^{k-1} G^2, \quad G^{2k+1}= (\alpha^2+\gamma^2)^{k} G.
\end{equation}
Therefore, 
\begin{widetext}
\begin{equation}
 e^{-\kappa G}={\mathbb I_3}+\Biggl(\frac{\cosh(\kappa\sqrt{\alpha^2+\gamma^2})-1}{\alpha^2+\gamma^2}\Biggr)G^2-
 \Biggl(\frac{\sinh(\kappa\sqrt{\alpha^2+\gamma^2})}{\sqrt{\alpha^2+\gamma^2}}\Biggr)G,\label{expo}
\end{equation}
\end{widetext}
where $\mathbb I_3$ denotes the three-dimensional unit matrix and $\kappa \in \mathbb C$. 

Taking into account the formula (\ref{expo}) we can show that  equation (\ref{gon}) reduces, after neglecting the $O(1/N)$ terms, to
\begin{widetext}
\begin{equation}
 g(t)= \frac{1}{[1+2\cosh(\beta \Theta)]^N}\Biggl\{\Bigl(2\cosh(\beta\Theta/2)\Bigr)^2\Biggl[\cos\biggl(\frac{J J_0 m t}{\Theta\sqrt{N}}\biggr)+i\frac{\Theta}{J}
 \sin\biggl(\frac{J J_0 m t}{\Theta\sqrt{N}}\biggr)\Biggr]^2-1\Biggr\}^N\label{gfun}.
\end{equation}
\end{widetext}
Expanding the cosine and the sine functions in Taylor series, we find that the modulus of $g$ is given by
\begin{widetext}
\begin{eqnarray}
 |g(t)|^2&=&\Biggl[\frac{\Bigl(2\cosh(\beta\Theta/2)\Bigr)^2-1}{1+2\cosh(\beta\Theta)}\Biggr]^{2N}\Biggl\{1+\frac{8m^2 J^2J_0^2 t^2\cosh(\beta\Theta/2)^2}
 {\bigl[1+2\cosh(\beta\Theta)\bigr]^2\Theta^2 N}\nonumber \\
 && \times\Biggr( [3+2\cosh(\beta\Theta) ]\frac{\Theta^2}{J^2}-1-2\cosh(\beta\Theta)\Biggr)+O(1/N^2)\Biggr\}^{N}.
\end{eqnarray}
\end{widetext}
We notice that the quantity within the square braces in the latter equation  is identically  equal to one. Hence, by taking the limit $N\to\infty$, it turns out that
\begin{eqnarray}
 \lim_{N\to \infty}|g(t)|^2 &=&\exp\Biggr\{-\frac{8 m^2 J_0^2 t^2\cosh(\beta\Theta/2)^2}
 {\bigl[1+2\cosh(\beta\Theta)\bigr]^2} \nonumber \\ 
 &\times & \Biggl([1+2\cosh(\beta\Theta)]\frac{J^2}{\Theta^2}-2\cosh(\beta\Theta)-3 \Biggr)\Biggr\}.
\end{eqnarray}
At first sight, the above formula  reveals that the evolution exhibits a Gaussian behavior, as expected. Nevertheless, a careful investigation shows that 
in order to ensure the decay of the modulus of $g$, the following condition has to be met
\begin{equation}
 \frac{\Theta^2}{J^2}<\frac{1+2\cosh(\beta\Theta)}{3+2\cosh(\beta\Theta)},\label{viol}
\end{equation}
otherwise the operator $\rho$ cannot be considered as a density matrix. This indicates  that the mean-field approximation breaks down if the above condition is violated.
For the numerical calculations, we have to take into account the condition~(\ref{viol}) together with the self-consistency equation (and the corresponding inequality for $w$)
\begin{equation}
 \frac{\Theta}{J}=\frac{4\sinh(\beta\Theta)}{1+2\cosh(\beta\Theta)}.
\end{equation}

Some comments are in order here. First of all, the latter equation is used to calculate numerically the order parameter $m$ for any particular values of the model parameters,
which is necessary for the subsequent discussion.  This means that the values of $m$ depend on the values of the other parameters. As the temperature increases, $m$ decreases
, and the  Gaussian decay becomes faster, and {\it vice versa}. Second, the time parameter will be given in units of the coupling constant $J_0$
which is very convenient as clearly noticed  by  inspecting  the  derivations presented above; to interpret the obtained results, 
it simply suffices to note that as $J_0$ increases, the Gaussian decay
becomes faster and {\it vice versa}. Third,
 as will be shown bellow, the condition~(\ref{viol}) determines only the ranges of the model parameters that yield a finite variation of the density matrix; it does not
 guarantee that the results of the mean-field approximation reproduce the exact dynamics. 
\begin{figure}[htb!]
 \centering
 \resizebox*{0.45\textwidth}{!}{\includegraphics{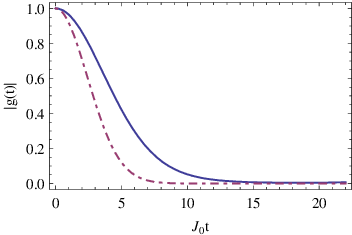}}
  \caption{(Color online)$|g(t)|$ as a function of the scaled time  $J_0 t$ for a transverse ising bath of $N=10000$ spin-1 particles with $T=2.52$ (solide line),
  and $T=2.54$ (dot-dashed line). The other parameters are $w=1$, $J=2$.
 These values yield the order parameters: $m=0.280$ for $T=2.52$, and $m=0.245$ for $T=2.54$. Note that we set $k_B=1$.}
 \label{fig3}
\end{figure}

In figure~\ref{fig3} we display the variation of $g(t)$ as a function of time for two different values of the temperature. It is clearly noticed  that the decay is Gaussian and gets faster as the temperature approaches $T_c$. This behavior is identical to that corresponding to
a spin bath that is composed of spin-$\frac{1}{2}$ particles. Nonetheless, it should be stressed that the direct comparison
of the decay in both cases is not possible, since the critical temperatures are not the same. Indeed, even when the baths have the same 
size, and the same coupling constants, the temperature ranges corresponding to the ordered phase when  $S=1$ and $S=\frac{1}{2}$ do not match. 
Moreover, it is evident  that the order parameters are different. 

\begin{figure}[htb!]
 {\centering
 \resizebox*{0.45\textwidth}{!}{\includegraphics{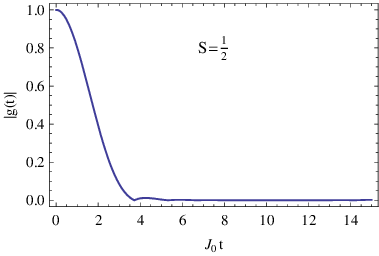}}
 \resizebox*{0.45\textwidth}{!}{\includegraphics{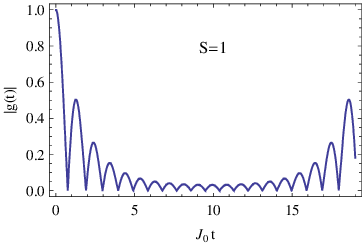}}
  \par}
  \caption{(Color online) $|g(t)|$ as a function of the scaled time $J_0 t$ when $w=0$ for  $S=\frac{1}{2}$ and $S=1$. Here   $N=10$ and
   $J= T$. Here  we take $k_B=1$. }
 \label{fig4}
\end{figure}

\begin{figure}[htb!]
 {\centering
  \resizebox*{0.45\textwidth}{!}{\includegraphics{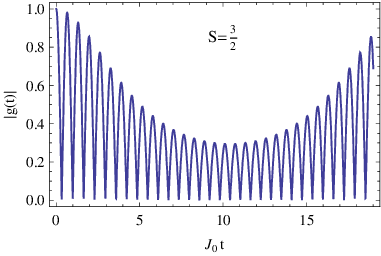}}
 \resizebox*{0.45\textwidth}{!}{\includegraphics{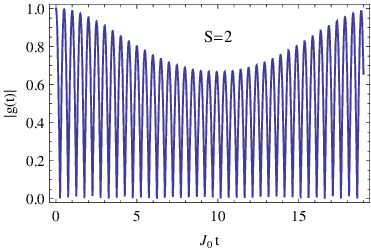}}
 \par}
  \caption{(Color online) The same as Fig.~\ref{fig4}  but for $S=\frac{3}{2}$ and $S=2$.}
 \label{fig5}
\end{figure}

\begin{figure}[ht]
 {\centering
 \resizebox*{0.45\textwidth}{!}{\includegraphics{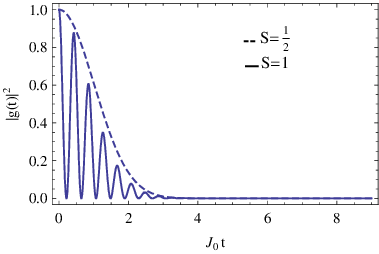}}
 \resizebox*{0.45\textwidth}{!}{\includegraphics{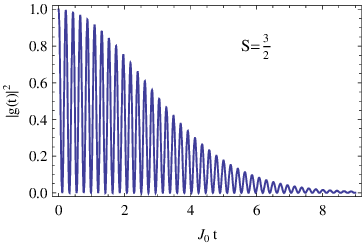}}
  \par}
  \caption{ (Color online) $|g(t)|^2$ as a function of the scaled time 
 $J_0 t$ when $w=0$ for some values of the spin $S$ with  $N=100$ and
   $J= T$ (we set $k_B=1$).}
 \label{fig6}
\end{figure}

Despite the breakdown of the mean-field approximation, one can draw some important conclusions  from the case of a vanishing transverse magnetic field, that is,
when $w=0$. This instance  can be studied exactly without resorting to any kind of approximations. Here the function $g(t)$ is given by

\begin{equation}
 g(t)=\frac{\sum\limits_j^{N S} \nu(j,N;S)\sum\limits_{\ell=-j}^j  \exp\Bigl\{i J_0 \ell t/\sqrt{N}+\beta J \ell^2/{N}\Bigr\}}{\sum\limits_j^{N S} 
 \nu(j,N;S)\sum\limits_{\ell=-j}^j
 \exp\Bigl\{\beta J \ell^2/{N}\Bigr\}},
\end{equation}
which can be evaluated numerically for arbitrary values of the coupling constants. 
An example of the variation in time of the modulus of the function $g(t)$ is depicted in figures~\ref{fig4} and~\ref{fig5} for several values of the spin $S$ when the size of the spin
bath is $N=10$; it can be seen that the loss of coherence  gets faster   a $S$ increases. The same result is found to be valid for any
higher value of the spin. This is identical to the short-time behavior associated with the dynamics of the qubit when it is coupled to the environment 
through Heisenberg $XY$
interactions. The other difference that can be inferred from the above figure is the presence of a oscillatory  periodic  collapse and  revival  
of the quantum coherence for finite size of the environment. These oscillations are mainly due to the competition between the two terms within the
exponential function in the numerator of the 
 expression of $g(t)$.  For a given $T$, they become more important as the size of the bath increases as shown in Figure~\ref{fig6}. The  latter behavior 
 is a mere  manifestation of the  
coupling between the qubit and the environmental constituents which is proportional to the magnitude of the spin. We notice also that 
 the aforementioned oscillations get  more and more  suppressed as the temperature increases.
This suppression is, however, quicker for small $S$ and is slower for larger $S$ which explains why they persist when $S>1/2$.

When  the temperature and the size of the spin bath
become sufficiently large, the oscillations completely disappear, and the Gaussian decay becomes dominant. Indeed, for example,  at high 
temperature we have
\begin{equation}
 g(t)\approx \biggl[\cos\Bigl(\frac{J_0 t}{2\sqrt{N}}\Bigr)\biggr]^N,
\end{equation}
 when $S=\frac{1}{2}$, whereas for $S=1$, 
\begin{equation}
 g(t)\approx 3^{-N}\biggl[1+2\cos\Bigl(\frac{J_0 t}{\sqrt{N}}\Bigr)\biggr]^N.
\end{equation}
 It follows that when $N$ becomes sufficiently large, the decay  is described by the laws:
\begin{equation}
 g(t)=\Biggl\{ \begin{array} {c}
 e^{-\frac{ J_0^2 t^2}{8}} \quad {\rm for} \quad S=\tfrac{1}{2}, \\
   e^{-\frac{ J_0^2 t^2}{6}} \quad {\rm for} \quad S=1,
   \end{array}
\end{equation}
which confirms the above observations.
\begin{figure}[ht]
 {\centering
 \resizebox*{0.45\textwidth}{!}{\includegraphics{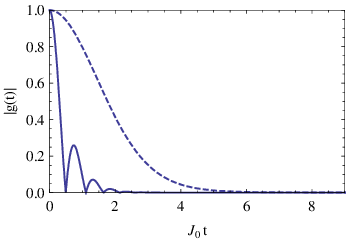}}
   \par}
  \caption{(Color online) Comparison between the exact (solid line)  and the mean-field (dashed line) solutions for $|g(t)|$ when $w=0$  for a spin bath of spin-$1$ particles.
 The parameters are  $N=100$, $J=3$  and $T=3.8$; the corresponding critical temperature is $T_c=4$, and the order parameter is $m=0.358$.}
 \label{fig7}
\end{figure}

For completeness, we have compared the results of the mean-field approximation with the exact one when $w=0$, see Fig.~\ref{fig7} for an example; it is clearly
seen that although  the mean-field solution produces a finite variation of the coherence, it fails to fairly reproduce
quantitatively the exact dynamics even when  condition~(\ref{viol}) is satisfied. 
To explain the reason for which the mean-field approximation
works for $S=\frac{1}{2}$ and fails for the other values, we note that the only  possible values of $S_z$ (which determines the order parameter $m$)
in the former case are $\pm\frac{1}{2}$,
that is they have the same magnitude. For
 larger values of $S$, this kind of uniqueness is lost, since $|S_z|=S,S-1\cdots 0(1/2)$, meaning that several  values of the spin  contribute asymmetrically with 
 different statistical weights to the
 value of the function $g(t)$. Let us stress, in the end, that the mean-field theory, as any other approximation, may or may not reproduce 
 the actual underlying physics of the studied system. It remains, however,  a useful tool to study, at least qualitatively, the dynamics in many cases.
\section{Conclusion}

Throughout this paper,  we have investigated several features of the decoherence  of a qubit,  due to its coupling  to a spin environment whose constituents are 
spin-$S$ particles. To achieve this aim, we have considered two particular types of the interaction between the central system and its surounding, namely,
Heisenberg $XY$ and
Ising interactions. In the first case, the form of the adopted Hamiltonian led  to the derivation of the degeneracy and the probability distribution of the total angular momentum,
obtained by summing the spin vectors of $N$ independent spin-$S$ particles. When the size of the spin bath is sufficiently large, the distribution turns out to be 
Gaussian, and we have 
determined its explicit analytical form. Because of the rescaling
of the coupling constants of the model, the obtained results are well-behaved when  $N$ is large. The long-time behavior has been studied  analytically
thanks to the probability distribution $P(j)$. The investigation shows that the asymptotic value of the coherence increases
with the magnitude of $S$, indicating that the recovery of the quantum interfernces is much better in environments for which $S$ is large. On 
the contrary, the short time decay is found to be inversely proportional to the spin.

In the case of a transverse Ising spin bath, we used the mean-field approximation to determine the ordered phase, and to identify the corresponding critical temperature;
the obtained results agree with those found by making use of the probability distribution of the total spin angular momentum.  
We have shown by analytical calculations that the mean-field approximation fails to provide us with acceptable physical results regarding the dynamics of the qubit. 
Indeed, we found that if the problem parameters do not satisfy certain additional conditions, then the off-diagonal elements of the reduced density matrix diverge as 
the time increases which is, clearly, unphysical. When the transverse magnetic field vanishes, it is shown that a collapse and a revival of the quantum coherence 
occurs if the size of the bath is not too large. As the size of the spin environment  as well as the temperature increase, these oscillations vanish, and the decay
is purely Gaussian. It is worthwhile mentioning at the end that the value of the critical temperature depends quadratically on $S$. This is quite important
from a practical point of view, 
since working at low temperatures requires cooling down the whole system; 
it is obvious that the greater $S$, the larger $T_c$ is, meaning that the  system may be exploited for eventual applications at relatively higher temperatures. 

The results reported in this study provide more insights into the effect of the spin environments on the dynamics of the quantum bits, which would contribute
to the understanding of the decoherence process, and to the quest for reliable experimental techniques that enable one to minimize this undesirable phenomena.

\acknowledgments
The author would like to thank the referee for the useful comments and suggestions.
\appendix{}
\section{Mean-field expression of $g(t)$ for $S=\frac{3}{2}$ , and $S=2$ \label{app1}}
\begin{widetext}
In this appendix we display the explicit form of the function $g$ defined in the main text by means of equation (\ref{gon}) when $S=\frac{3}{2}$ , and $S=2$.
\subsection{$S=\frac{3}{2}$}
In this case,
\begin{eqnarray}
 g(t)&=&\frac{\cosh(\beta\Theta/2)^N}{\Bigl(\cosh(3\beta\Theta/2)+\cosh(\beta\Theta/2)\Bigr)^N} \Biggl[\cos\biggl(\frac{J J_0 m t}{\Theta\sqrt{N}}\biggr)+i\frac{\Theta}{J}
 \sin\biggl(\frac{J J_0 m t}{\Theta\sqrt{N}}\biggr)\Biggr]^N\nonumber \\
 &\times &\Biggl \{\Bigl(2\cosh(\beta\Theta/2)\Bigr)^2\Biggl[\cos\biggl(\frac{J J_0 m t}{\Theta\sqrt{N}}\biggr)+i\frac{\Theta}{J}
 \sin\biggl(\frac{J J_0 m t}{\Theta\sqrt{N}}\biggr)\Biggr]^2-2\Biggr\}^N.
 \end{eqnarray}
From the above expression, we find that the modulus of $g$ is given by
\begin{eqnarray}
 |g(t)|^2&=&\Biggl\{1+\frac{m^2 J^2J_0^2 t^2}
 {2\bigl[\cosh(\beta\Theta)\bigr]^2\Theta^2 N} \Biggr( \bigl[11+12\cosh(\beta\Theta)+3\cosh(2\beta\Theta) \bigr]\frac{\Theta^2}{J^2}
 \nonumber \\
 &-&\bigl[3+4\cosh(\beta\Theta)+3\cosh(2\beta\Theta) \bigr]\Biggr)+O(1/N^2)\Biggr\}^{N}.
\end{eqnarray}

 It fillows that
 \begin{eqnarray}
 \lim_{N\to \infty}|g(t)|^2 &=&\exp\Biggr\{-\frac{m^2 J_0^2 t^2}
 {2\bigl[\cosh(\beta\Theta)\bigr]^2 }  \Biggl(\bigl[3+4\cosh(\beta\Theta)+3\cosh(2\beta\Theta) \bigr]\frac{J^2}{\Theta^2}
 \nonumber \\ 
 && -\bigl[11+12\cosh(\beta\Theta)+3\cosh(2\beta\Theta) \bigr] \Biggr)\Biggr\}.
\end{eqnarray} 
 To ensure the Gaussian decay we should have
 \begin{equation}
  \frac{\Theta^2}{J^2}<\frac{3+4\cosh(\beta\Theta)+3\cosh(2\beta\Theta)}{11+12\cosh(\beta\Theta)+3\cosh(2\beta\Theta)}.
 \end{equation}
\subsection{$S=2$}
Here the function $g$ reads
\begin{eqnarray}
 g(t)&=&\Biggl\{-2+\Bigl(2\cosh(\beta\Theta/2)\Bigr)^4\Biggl[\cos\biggl(\frac{J J_0 m t}{\Theta\sqrt{N}}\biggr)+i\frac{\Theta}{J}
 \sin\biggl(\frac{J J_0 m t}{\Theta\sqrt{N}}\biggr)\Biggr]^4\nonumber \\
 &-&3\Biggl(-1+ \Bigl(2\cosh(\beta\Theta/2)\Bigr)^2\Biggl[\cos\biggl(\frac{J J_0 m t}{\Theta\sqrt{N}}\biggr)+i\frac{\Theta}{J}
 \sin\biggl(\frac{J J_0 m t}{\Theta\sqrt{N}}\biggr)\Biggr]^2\Biggr)  \Biggr\}^N\nonumber\\
 &\times& \frac{1}{\Bigl(1+2\cosh(\beta \Theta)+2\cosh(2\beta\Theta)\Bigr)^N}.
\end{eqnarray}
This yields
\begin{eqnarray}
 |g(t)|^2&=&\Biggl\{1+\frac{8m^2 J^2J_0^2 t^2 (\cosh(\beta\Theta/2))^2}
 {\bigl[1+2\cosh(\beta\Theta)+2\cosh(2\beta\Theta)\bigr]^2\Theta^2 N}\nonumber \\ &\times & \Biggr( \bigl[31+42\cosh(\beta\Theta)+18\cosh(2\beta\Theta)
  +4 \cosh(3\beta\Theta) \bigr]\frac{\Theta^2}{J^2}\nonumber \\&-&\bigl[5+10\cosh(\beta\Theta)+6\cosh(2\beta\Theta)+4
\cosh(3\beta\Theta) \bigr]\Biggr)+O(1/N^2)\Biggr\}^{N}.
\end{eqnarray}
Therefore:
\begin{eqnarray}
\lim_{N\to \infty}|g(t)|^2 &=&\exp\Biggl\{-\frac{8m^2 J_0^2 t^2 (\cosh(\beta\Theta/2))^2}
 {\bigl[1+2\cosh(\beta\Theta)+2\cosh(2\beta\Theta)\bigr]^2}\nonumber \\ &\times & \Biggr( \bigl[5+10\cosh(\beta\Theta)+6\cosh(2\beta\Theta)+4
\cosh(3\beta\Theta) \bigr]\frac{J^2}{\Theta^2}\nonumber \\&-&\bigl[ 31+42\cosh(\beta\Theta)+18\cosh(2\beta\Theta)
  +4 \cosh(3\beta\Theta)\bigr]\Biggr)\Biggr\},
   \end{eqnarray}
 meaning that
 \begin{equation}
  \frac{\Theta^2}{J^2}<\frac{5+10\cosh(\beta\Theta)+6\cosh(2\beta\Theta)+4
\cosh(3\beta\Theta)}{ 31+42\cosh(\beta\Theta)+18\cosh(2\beta\Theta)
  +4 \cosh(3\beta\Theta)}.
 \end{equation}
\end{widetext}
   \section{Evolution of the bosonic system defined by equations~(\ref{bose1}) and (\ref{bose2})\label{app2}}
\begin{figure}[htb!]
 {\centering
 \resizebox*{0.65\textwidth}{!}{\includegraphics{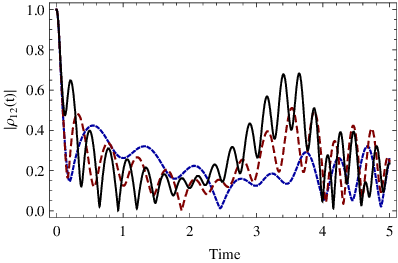}}
   \par}
  \caption{(Color online) Evolution in time of the modulus of $\rho_{12}$ given by Eq.~\ref{bose3} for: $S=5$ (dotted line), $S=8$ (dashed line), and $S=12$ (solid line).
  The other parameters  are $g=1$, $\beta=0.01$, $\mu=3$  and $\alpha=0.5$. Note that $\rho_{12}(0)$ is normalized to unity for convenience, and the time is given in unit of 
  $\alpha^{-1}$. }
 \label{fig8}
\end{figure}

 The aim of this appebdix is to investigate the dynamics of the qubit using the  bosonic model 
   \begin{eqnarray}
   H_S&=&\mu\sigma_z,\\
  H_B&=&2gSB^\dag B,\\
 H_{SB}&=& 2\alpha\sqrt{2S}(\sigma_- B^\dag+\sigma_+ B),
\end{eqnarray}
which has been obtained via the Holstein-Primakoff transformation. The time evolution operator corresponding to the above system can be
determined using the Schr\"odinger equation:
\begin{equation}
 i\frac{dU(t)}{dt}=(H_S+H_{SB}+H_B)U(t).
\end{equation}
In the standard basis of $\mathbb C$, we find that the components of the above operator are given by
\begin{widetext}
\begin{eqnarray}
 U_{11}(t)&=&e^{-4ig tS(\hat n-\frac{1}{2})}\Bigl[\cos(t M_1)-\frac{i(gS-\mu)}{M_1}\sin(t M_1)\Bigr],\\
 U_{22}(t)&=&e^{-4ig t S(\hat n+\frac{1}{2})}\Bigl[\cos(t M_2)+\frac{i(gS-\mu)}{M_2}\sin(t M_2)\Bigr],\\
 U_{12}(t)&=&U_{21}^\dag(t)=-4i\alpha\sqrt{2S} B^\dag e^{-i(\mu+2gS\hat n)t/2} \sin(t M_2)/M_2,
\end{eqnarray}
\end{widetext}
where 
\begin{eqnarray}
 \hat n&=&B^\dag B,\\
 M_1&=&\sqrt{(gS-\mu)^2+8\alpha^2S\hat n},\\
 M_2 &=&\sqrt{(gS-\mu)^2+8\alpha^2S(\hat n+1)}.
\end{eqnarray}
The partition function of the bath, which is assumed at inverse temperature $\beta$ can easily be calculated as:
\begin{equation}
 Z=\sum_{n=0}^\infty e^{-2Sg\beta n}=\frac{e^{2S\beta}}{e^{2S\beta}-1}.
\end{equation}
Hence, assuming a factorized initial state, the evolution in time of the coherence in this approximation is described by
\begin{widetext}
\begin{equation}
 \rho_{12}(t)=\rho_{12}(0)e^{-4i g S t}(1-e^{-2gS\beta})\sum\limits_{n=0}^\infty\Bigl[e^{-2gS\beta n}\langle n| U_{11}(t)U_{22}^*(t)|n \rangle\Bigr]\label{bose3}.
\end{equation}
\end{widetext}
For convenience, we  displayed in figure~\ref{fig8} an example of the dependence on  time of the modulus of $\rho_{12}$; it can be seen that the latter does not assume constant values
at long times, but rather oscillates  as the time increases. This is due to the fact
that  all the terms that depended on the spin under the square root in the Holstein-Primakoff transformation have been omitted. We can however assure that  at long times,
 $\rho_{12}$ assumes larger values as the magnitude of the spin increases. At short time, the decay is almost independent of the value of $S$. 

\end{document}